# NOTE:

## This manuscript is superseded by a substantially revised and extended version, “Hybrid Reasoning Systems That Prioritize and Enhance Human Intelligence”, published in Int. J. Human–Computer Interaction (2026

## Please cite the published version.

# Developing “Full Stack” Augmented Human Intelligence

SEAN KOON IS A PHYSICIAN AND INNOVATION ADVISOR WITH THE SOUTHERN CALIFORNIA PERMANENTE MEDICAL GROUP IN FONTANA, CALIFORNIA, USA

The idea of augmented or hybrid intelligence offers a compelling vision for combining human and AI capabilities, especially in tasks where human wisdom, expertise, or common sense are essential. Unfortunately, human reasoning can be flawed and shortsighted, resulting in adverse individual impacts or even long-term societal consequences. While strong efforts are being made to develop and optimize the AI aspect of hybrid reasoning, the real urgency lies in fostering wiser and more intelligent human participation. This paper proposes the development of generative-AI-based tools that are designed to enhance the full spectrum of what it means to reason well, from reflection on values to in-depth analysis of data. By enhancing critical thinking, ingenuity, expertise, and even wisdom, such tools could be essential towards helping people to address the challenges of our emerging world.

CCS CONCEPTS •Computing methodologies~Artificial intelligence •Human-centered computing~Human computer interaction (HCI)~HCI theory, concepts and models

**Additional Keywords and Phrases:** augmented intelligence, hybrid reasoning, expertise, critical thinking

## 1 INTRODUCTION

AI tools, particularly advanced Generative AI (GenAI) models, are poised to create exponential change, demonstrating remarkable reasoning capabilities previously thought exclusive to humans. However, this raises concerns about adverse impacts as well as concerns about human agency and autonomy[27, 34]. Many promote an "augmented intelligence" approach, acknowledging that humans and computers excel in different areas, which, when combined, can achieve superior outcomes. Central to this vision is human contribution and control, ensuring that wisdom, tacit knowledge, expertise, and common sense inform problem-solving[22, 34].

A potential flaw in this idea of human participation and control is that we are not reliable reasoners ourselves. We are subject to an exhaustive number of cognitive biases and reasoning fallacies[1] and have emotions that interact in complex ways with our decisionmaking[20]. We pursue short-term solutions, leading to long-term issues like climate change, bureaucratic traps, healthcare disparities, and political polarization[17, 36]. Human reasoning tends to be "hasty, narrow, fuzzy, and sprawling[29], weaknesses that we are likely to contribute to hybrid intelligence. While AI cannot be trusted to make good decisions for us, often, neither can we.

There is an urgency in that AI-equipped humans can now create change at an exponential rate, but without any similar increase in wisdom about how to shape or adapt to these changes. This paper proposes a science of prompting humans, stimulating them to be wise, to exercise critical thinking skills, to think "outside the box", and to exercise greater expertise in the tasks that they face.

## 2 HUMAN-CENTERED HYBRID REASONING

The hybrid reasoning discussed here is "human-centered" in the sense that it takes a primary interest in optimizing the user's intellectual performance vs. the optimization of AI performance. The rationale for this approach is that there are highly consequential AI-supported tasks that require not only human input, but in some cases, the best of our reasoning. The conceptual model of the Data, Information, Knowledge, Wisdom(DIKW) pyramid is illustrative here, describing categories of reasoning elements and the transformations that occur between them[32] (figure 1). It also generally represents elements we expect humans to be better at (the top of the stack) vs. AI (the bottom of the stack). The opportunity here is to is to develop tools that have two functions, both enhancing AI-supported analysis as well as stimulating and scaffolding higher level human reasoning. This bidirectional action is discussed in section 8, where these two functions can be orchestrated to support a human-AI reasoning dialogue. However, the initial question to consider is whether (and how) we might augment human intelligence in a very literal sense, i.e. improve the quality of what happens within the human mind.

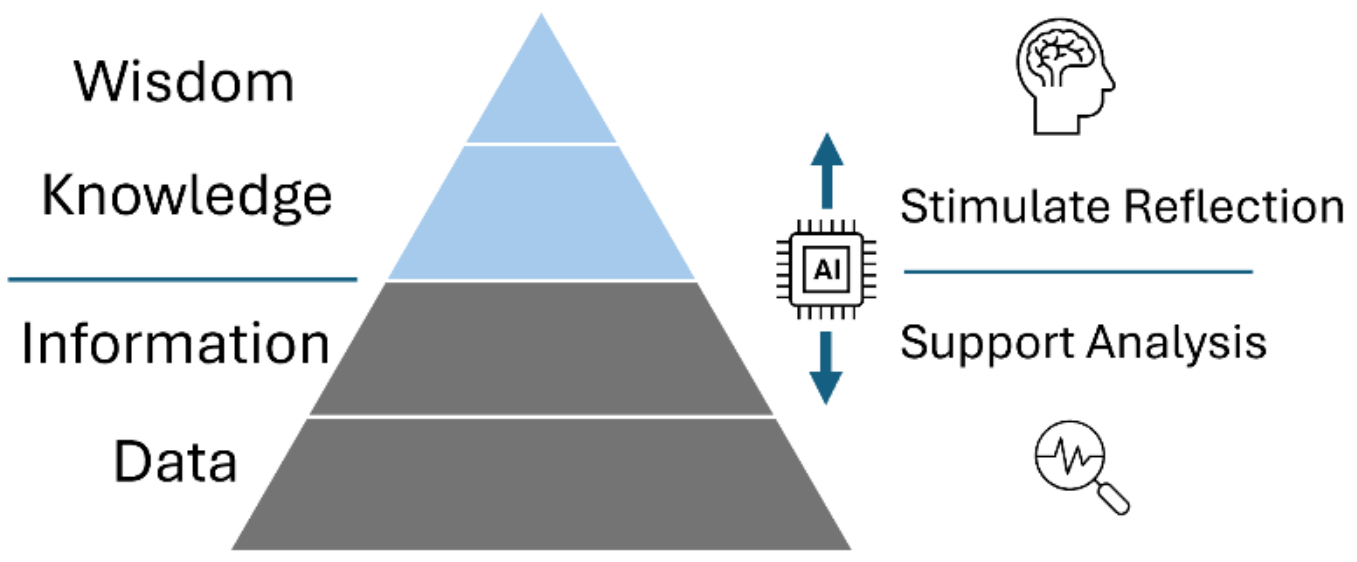

Figure 1. Hybrid reasoning and the Data, Information, Knowledge, Wisdom (DIKW) pyramid

## 3 ENHANCING HUMAN REASONING WITH GENERATIVE AI

Efforts to improve reasoning range from the teachings of ancient philosophy to more modern disciplines such as statistics or medical ethics. For example, there is a robust and well-established curriculum on "Critical Thinking" which aims to enhance skills in deductive reasoning, identifying cognitive biases, group decisionmaking, hypothesis testing, probabilistic decisionmaking, complex problem solving, etc.[18].

The impact of teaching critical thinking is limited for a variety of reasons. For one, these courses show only moderate or inconsistent improvements[31]. Also, these skills are potentially learned years before they are needed. Perhaps most importantly, these skills must be translated or "transferred" to novel problems which may not resemble the examples with which they were taught[26]. Also, most people do not take such courses, creating a disparity of access. A more ideal solution would involve tools that can not only bring such skills into the present moment but can also translate the critical thinking concepts to fit the immediate real-world problem and context.

### 3.1 Leveraging the synthetic, language-based, ubiquitous nature of Generative AI

GenAI models may represent a new solution for this old problem, primarily due to their striking if sometimes comical synthetic capabilities. One might prompt ChatGPT to generate song lyrics about quantum physics, while adhering to a specific structure of verses and choruses, and in the style of the rock band Metallica. With unprecedented capability, the GPT manages to synthesize diverse conceptual content including domain expertise (quantum physics), a personal style (Metallica), and a scaffold (the song structure).

This frivolous example is not far from much more useful possibilities, such as giving GenAI a RAG (Retrieval Augmented Generation) resource on ethical principles (interpretive framework) and having it "propose a set of ethical considerations" (a reasoning strategy) that might pertain to a challenging medical or political decision (a domain context), while considering viewpoints of potential stakeholders (personas).

Another crucial feature is GenAI's facility with language and dialogue. Through human-AI dialogue, we could leverage a variety of common reasoning strategies such as critique, argumentation, deliberation, hypothesizing, brainstorming, and much more. Argumentation is an especially appealing way to engage and enhance human reasoning[10]. For example, GenAI could challenge users with "other-side views", a crucial aspect of both critical thinking and wisdom[31].

Finally, GenAI is increasingly ubiquitous, creating a new opportunity for applying effective reasoning skills at the precise moment that they are needed.

Fortunately, strategies to enhance human reasoning have been well-explored in the domains of cognitive psychology, learning theory, critical thinking courses, innovation techniques, etc. These sources should be an essential foundation for hybrid reasoning strategies. Four potential categories of enhancements are briefly discussed here: critical thinking, innovation, wisdom, and expertise.

## 4 ENHANCED CRITICAL THINKING

Critical thinking(CT), as the term implies, employs systematic methods to critique, analyze, and challenge the credibility of information. Significantly, this critical thinking is also self-corrective in that it can be directed towards our own thoughts, assumptions, or biases[10].

The need for critical thinking spans the full spectrum of thinkers. Critical thinking approaches are not inherently possessed by people, even those with high intelligence or expertise[1, 29]. The roles of business leaders, policymakers, judges, physicians, and other professionals who influence human lives warrant the support of CT tools. However, CT is also essential in everyday reasoning tasks[12], where individuals face decisions including large purchases, consequential healthcare decisions, voting choices and much more. New and powerful campaigns of persuasion and misinformation make CT more essential than ever before. As Nickerson[26] suggests, "A thinking citizenry is seen by many as essential to the preservation of a democratic way of life."

There is an incredibly low threshold for introducing CT to tasks with GenAI. Simply adding a button to a news browser or word processor that prompts GenAI to "use critical thinking principles to analyze this text for reasoning fallacies" could perform usefully without any added RAG. For other tasks, the prompts might leverage RAG enhancements grounded in established CT concepts that are readily available in a general textbook[18]. While simple interventions such as these may not be technically impressive, they serve a powerful purpose simply by signaling that default reasoning can be imperfect and by describing good CT practices.

More advanced CT tools could support a range of reasoning tasks from analysis and sensemaking to decisionmaking and problem-solving. They might assist in both generating hypotheses and evaluating their plausibility[21, 28]. In decisionmaking tasks, our tools could help us identify our values, outline pros and cons, or nudge us to compare options from multiple perspectives. CT tools could assist people in statistical reasoning tasks for which we are not innately adept, such as predicting likelihoods, or calculated expected the expected utility of decision options[13].

Deliberation or argumentation-based interactions will likely be "go-to" designs for hybrid reasoning. Also, GenAI might be used to provide an external analysis to human arguments, such as those found in a white paper or even a committee meeting in progress, etc.[24]. AI tools could create a visual diagram of arguments, perhaps using the Toulmin model, mapping out claims, grounds, rebuttals, etc.[11, 35]. Helping people to externalize their reasoning may also create longer lasting enhancement of reasoning skills[35]. The promising potential of these tools lies in their capability to facilitate in-task learning of CT skills, which may be transferable to other contexts where AI is not employed.

## 5 ENHANCED INNOVATIVE THINKING

Creativity, innovation or "out of the box" thinking is crucial for solving new challenges as well as old, intractable ones. By enhancing creative thinking we could "expand the range of thoughts we can think"[7]. The synthetic abilities of GenAI are particularly suited to innovation and could be used in a variety of ways to "increase the quantity, quality, and diversity of ideas and to create more value in their innovation processes—at a very low cost"[4]. Even GenAI's "hallucinations" or nonsense outputs may be more of a feature than a bug; the primary goal of brainstorming is to prompt human innovative thinking, and outlandish ideas often lead to very useful ones. The literature contains many innovation techniques and resources that might be applied with GenAI. Edward de Bono, for example, has created exhaustive strategies and frameworks to stimulate innovative thinking and problem solving[8].

The TRIZ framework represents one example of a highly developed innovation tool. This framework originated from a review of thousands of patents, identifying the pivotal innovation in each invention. This review resulted in a collection of primary innovation elements such as "use pulsed action"; "use beforehand cushioning"; "merge objects or functions", etc., that could be applied experimentally to an engineering problem. With GenAI's ability to apply abstract strategies to concrete problems, these types of frameworks represent an appealing opportunity for hybrid reasoning in the innovation space[25, 30].

Other frameworks could prove useful across a spectrum of innovation tasks such as opportunity analysis, needs identification, user characterization, idea generation and more[2, 33]. Furthermore, after an idea is generated, GenAI could facilitate rapid prototyping and testing, thus accelerating the design cycle[4].

## 6 ENHANCED EXPERTISE

We seek experts for their ability to create a desired outcome using their "special skill, knowledge, or judgment"[19]. Not only do experts have more knowledge but they "organize it differently"[12] with an understanding of the key "cues" or reliable shortcuts that can rapidly lead to next steps[5, 14, 20].

The prevailing model for attaining expertise usually involves extensive periods of study and apprenticeship[9]. However, even well-trained experts may have knowledge gaps from a lack of specific experience or from skill decay. Crucially, the need to rapidly acquire "just-in-time" expertise becomes more urgent as our work becomes more cognitive and as technological advances increase the rate of change[38].

Also, outside the professional realm, most people must become "novice experts" in choices such as large purchases, medical decisions, investments, voting, etc. Often, people do not have access to experts, there may be no expert, or the expert may be biased because they are also promoting the product or idea for which expertise is needed. In short, our model for expertise is not future-proof, it is subject to skill gaps and decay, and it makes expertise inaccessible to many.

### 6.1 Resourcing GenAI tools for enhanced human expertise

GenAI tools have rapidly become impressive "answer machines" for general users and show potential in expert domains[15]. However, issues of reliability and bias are significant concerns for high-impact decisions. From a human-centered hybrid reasoning perspective, a better goal would be to preserve human agency by making AI involvement "preconclusive" in that it doesn't supply solutions but rather it provides pertinent information along with potential reasoning strategies. Furthermore, it would be deliberative, where instead of generating a solution, our tools would facilitate a dialogue between human and machine regarding options, exploring outcomes, testing hypotheses, evaluating tradeoffs, etc.[23].

#### *6.1.1Expert information and strategies*

Tools that enhance expertise should be able to answer the following questions: 1. What information is relevant and reliable for this task? and 2. What are the efficient and practical ways to apply this information to this problem? These tools would be augmented with specialized and reliable information(RAG), which could be obtained from textbooks, ontologies or knowledge graphs, workflows, manuals, guidelines, scientific reports, diagrams, and schematics, etc.

Beyond expert information, expertise requires reasoning strategies for apply information efficiently towards an overall goal. How might we use AI to model the cognitive work of experts in ways that could enhance the reasoning of those with less expertise? How can GenAI-based tools provide scaffolding to help novices and experts to organize their problem solving or decisionmaking? How might large case repositories of expert decisions be used to support case-based reasoning processes?[6], etc.

#### *6.1.2Using GenAI to enhance expertise for novel problems*

In a world of accelerating change, there will be new questions and problems for which there are no well-established reasoning strategies. We might create general models of reasoning tasks such as sensemaking, hypothesis testing[23], weighing decisional tradeoffs, etc., and employ GenAI to help users to apply general reasoning strategies to novel problems. Again, the incredible ability of GenAI to apply abstract concepts to specific contexts could be essential in helping users in approaching future problems "like an expert" in situations where no expertise is yet established.

## 7 ENHANCED WISDOM

Wisdom may seem an unlikely target for computer intervention. Rather, we may see it as one of those key human contributions that we hope to introduce into our "human-in-the-loop" designs. By involving people, we hope that human impacts, broader ramifications, and future possibilities are considered. Nevertheless, attention is a limited resource in the face of complexity and excessive information. We may lack the motivation, ability, or mindfulness to reason beyond the immediate demands of our tasks unless we are somehow prompted and assisted in doing so.

What is wisdom and how could it be enhanced? With thousands of years of wisdom literature, wisdom is difficult to define both precisely and comprehensively. However, in a vast simplification, we can suggest that wisdom has a "primary concern with values"[3]. It is concerned about impacts beyond short-term goals (longer vision) but also has a broader, adaptive vision, that balances individual and societal benefits[36]. Wisdom also has an aspect of reflectiveness or "metacognition" that is characterized by humility, i.e. an openness to pursue truth beyond current beliefs or assumptions[16]. Thus, AI tools that specifically help decisionmakers adopt a longer term vision, consider broader human impacts, and reflect on higher values could conceivably enhance human wisdom.

For long-term vision, we might aim to use AI to understand mechanisms by which our short term targets are depleting resources for the whole group in the long term (i.e. the "tragedy of the commons" phenomenon). In the reverse, our tools could identify specific collective and cooperative behaviors that might generate benefits for all.

AI tools could assist in exploring broader impacts and higher values in our decisions. For example, an investment tool that compares the financial performance of various companies has a narrow value focus. One that also predicts the ecological or human rights impacts of a company has a wider, potentially wiser lens of analysis. GenAI personas of different stakeholders or groups might be used to broaden the range of perspectives that are considered. Tools that present viable "other-side" views to an issue could enable broader thinking. Tools that enhance reflectiveness or metacognition could help individuals to gain a deeper understanding of their own reasoning patterns or biases. Likewise, summarization tools could assist committees in evaluating their reasoning and priorities by analyzing the meeting content in relation to the stated agenda, organizational values, or established goals.

## 8 ENVISIONING "FULL STACK" INTERACTIVE REASONING SUPPORT

Earlier this paper referenced the DIKW pyramid, emphasizing the use of GenAI to stimulate and enhance the "upper stack" aspects of reasoning such as wisdom and critical thinking. Yet, the real power of hybrid reasoning occurs when we integrate this AI-enhanced upper-stack reflectiveness with AI-enabled lower-stack exploration such as data analysis, prediction, simulation, etc. (figure 2). By AI-enabled we mean that the GenAI, when given tasks by the user, can access a set of tools for research and analysis.

These two elements permit a reciprocal, iterative dialogue whereby the user and AI can explore a problem space by prompting and querying each other. AI could help to stimulate and structure the reasoning of the user who, after reflection, could direct the AI to perform an analysis, test a hypothesis, make a prediction, etc.

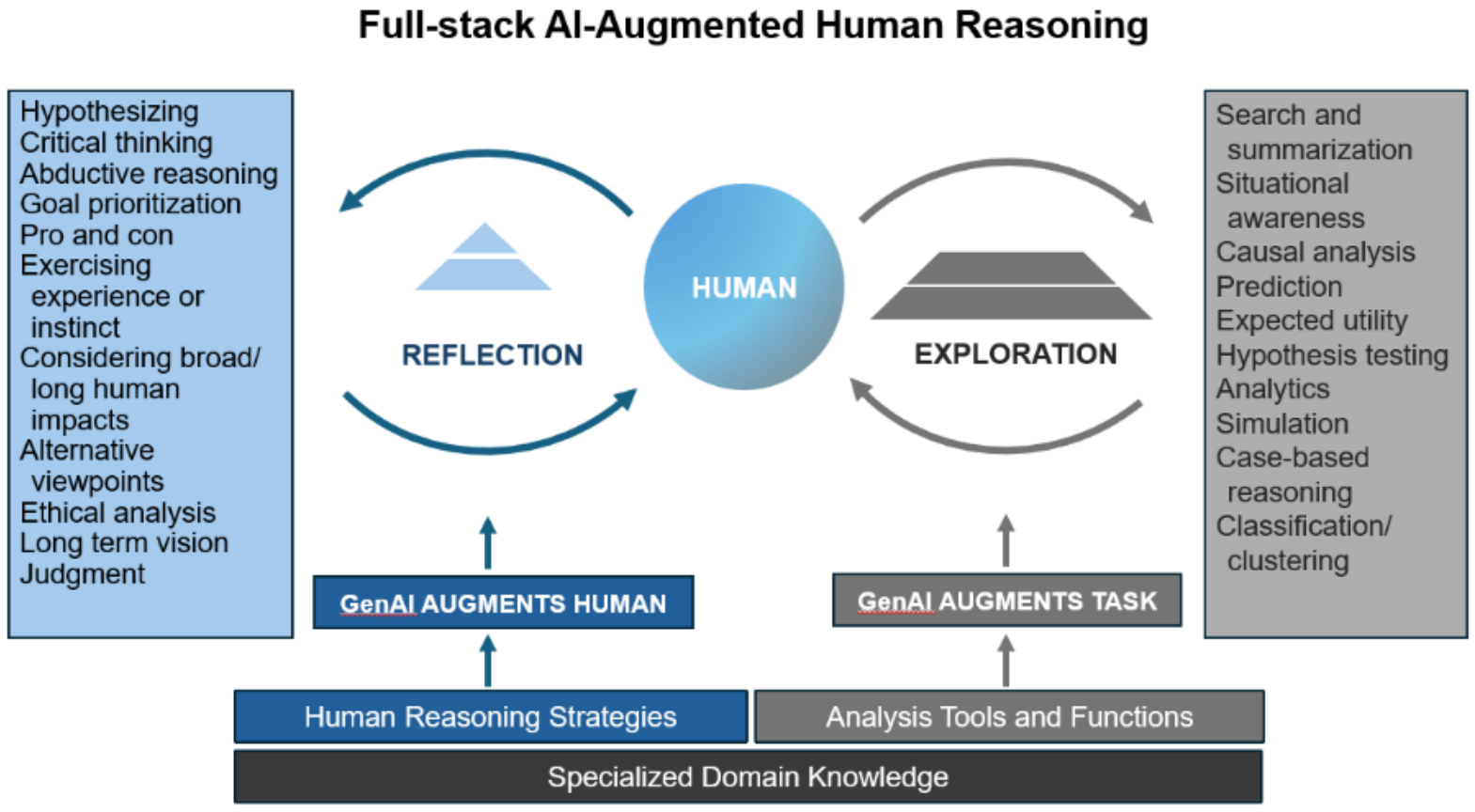

Figure 2. Full-stack AI-augmented human reasoning

The goal of "full stack" reasoning support is to create an end-to-end connection between abstract human values and the vast, granular data of our modern world. For a low-tech example of full-stack reasoning, we might consider an interaction between a personal finance consultant and their client. These experts assist their clients in making wise, values-based decisions by helping them to reflect upon a variety of personal values and goals that they might not have considered. The expert can then carry these goals through each layer of the pyramid, applying their expertise and knowledge about taxes or investments, and considering data regarding market trends, expense trends, or some other prediction. These insights can then translate up through the pyramid and lead to adjustments in the client's high-level plans. This process can iterate until personal values are harmonized in a practical way with the financial realities of the individual and of the world.

## 9 BUILDING A SCIENCE OF REASONING SUPPORT

An emphasis on human reasoning creates a very divergent set of research questions to pursue (figure 3) [figure 3 about here or after this paragraph, may require full page]. There is a need to revisit prior research on effective reasoning and begin to build a translational science of applying these learnings towards the human-AI context. When taking a perspective of "humans-in tasks", this is a need to be able to characterize and model the reasoning that occurs within a task or domain, including existing risks related to complexity, bias, or fallacious reasoning. To build reasoning support, the larger tasks of "decisionmaking" or "problem-solving" would need to be broken down into enhanceable subtasks such as sensemaking, hypothesis testing, causal analysis, etc., for which microtools might be created. Realistically the front-end experience of working with "an AI" will involve orchestrating multiple if not many AI's in the background[39].

Dialogue is another important challenge. Currently our “dialogue” with GenAI involves trying to ask good questions to achieve optimal GenAI outputs. In contrast, optimizing human reasoning requires a 2-layer strategy of prompting the AI to then prompt the human. These exchanges require an overall dialogue orchestrator that moderates any deliberation, argumentation, exploration or analyses, and helps to organize the results of these exchanges. There are also important challenges around visualization as well, as this “dialogue” will often be multimodal, involving documents, dashboards, maps, schematics, lines of code, etc., with interfaces that support both direct manipulation of these artifacts as well as a chat function to discuss them.

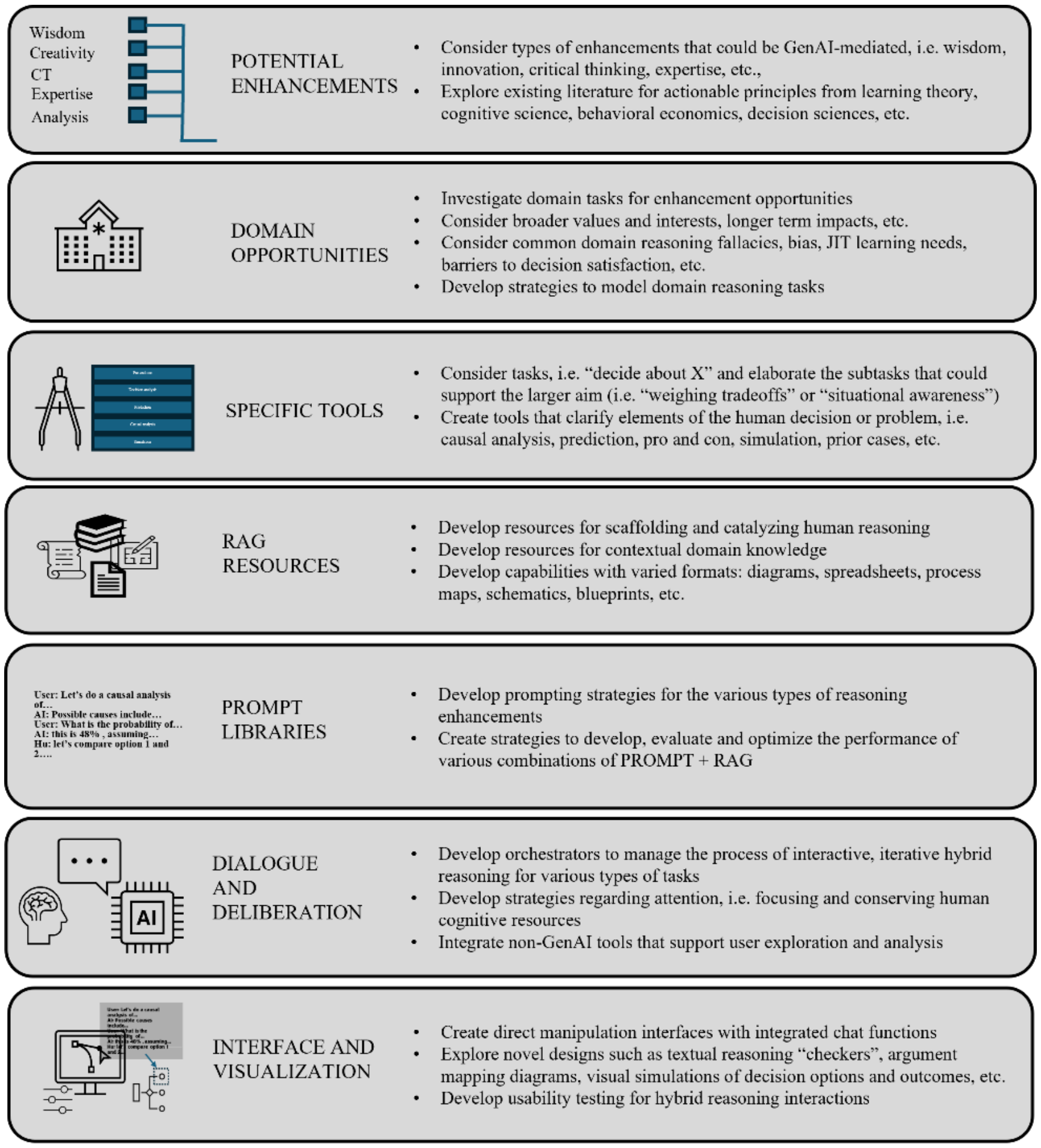

Figure 3. Overview of potential areas of research and development for reasoning enhancement tools.

Reasoning tools might be especially useful for human-impactful problems involving complex information, high degrees of uncertainty, or where multiple potentially competing aims or values apply. Such tools might be an essential asset in geopolitics, civic planning, healthcare, defense, lawmaking, and many other domains whether professional or personal. Perhaps we might even gain traction on important global goals, such as the UN’s Sustainable Development Goals regarding global hunger, poverty, responsible consumption and production, clean energy, sustainable cities, etc.[37]. As intelligence researcher Robert Sternberg suggests, “What is intelligence for, if not adaptation?”[36].

## 10 DISCUSSION

The idea of enhancing human reasoning might be an appealing one but is far from inevitable. There is a great momentum to optimize AI as a general intelligence, increasing the range of human-tasks that it can do. Other efforts are productivity focused, increasing the number and quality of things that we can build, publish, or sell to meet growing demands or reduce costs. While in theory, investing in human “wisdom”, “reflectiveness”, “expertise” or “critical thinking” might benefit humanity, the objectives are currently poorly defined and they face competition from more immediate and quantifiable short-term optimizations of performance metrics, quarterly financials, election terms, and other factors.

In all of this, there remains a question of the will to reason. In a story that is both humorous and disheartening, critical thinking educator Diane Halpern describes her preparation to speak before the Congressional Committee on Science. In an ironic interaction, she was coached by the congressional staff to avoid using statistics, stating that persuasive anecdotes would be more appropriate for this group of leaders[18].

It is easier to reason superficially, and we may have strong motivations to maintain old beliefs, to ignore undesirable facts, or assume our reasoning is sound. As we face complex new challenges, we might find our decisions to be burdensome and unfamiliar, creating a temptation to defer to self-described experts whether human or artificial.

If there was a single accomplishment that we could achieve with GenAI, it would be to lower the threshold for powerful and wise human reasoning. Well-designed tools might help arrange and simplify complex problems so that reasoning becomes more manageable and rewarding. We might even hope for a cognitive “herd immunity” such that after many people had been coached by reasoning tools, that there would be a more general awareness of poor arguments and misleading persuasions. At some point, such misleading content could even be seen as worthy of disdain rather than getting the attention and acceptance that much misinformation receives today.

There has always been some fear that intelligent robots or strong AI will cause a dystopian future. The great irony, indeed, important possibility, is that human-centered AI tools could facilitate and popularize the kinds of reasoning that could prevent not only AI, but also human narrowmindedness from creating such dystopias.